\documentstyle[a4,11pt]{article}
\begin{document}

\begin{titlepage}
\vskip 2cm
\begin{flushright}
Preprint CNLP-1994-04
\end{flushright}
\vskip 2cm
\begin{center}
{\bf THE GAUGE EQUIVALENCE OF THE ZAKHAROV  EQUATIONS AND
(2+1)-DIMENSIONAL CONTINUOUS HEISENBERG FERROMAGNETIC
MODELS}\footnote{Preprint
CNLP-1994-04. Alma-Ata.1994 }
\end{center}
\vskip 2cm
\begin{center}
{\bf R. Myrzakulov }
\end{center}

\vskip 1cm
Centre for Nonlinear Problems, PO Box 30, 480035, Alma-Ata-35, Kazakhstan

E-mail: cnlpmyra@satsun.sci.kz

\vskip 1cm

\begin{abstract}

The gauge equivalence between the (2+1)-dimensional
Zakharov equation and
(2+1)-dimensional integrable
continuous Heisenberg ferromagnetic model is established.
Also their integrable
reductions are shown explicitly.

\end{abstract}


\end{titlepage}

\setcounter{page}{1}
\newpage
\Large

The concepts of gauge equivalence between completely integrable equations
plays important role in the theory of solitons[1,2]. In the (2+1)-dimensions
such equivalence have been constructed recently for the Davey-Stewartson and
Ishimori equations[3], for the some Myrzakulov and nonlinear Schr\"odinger
type equations and so on[4-7].  In this Letter we wish find the gauge
equivalent counterparts of the some (2+1)-dimensional integrable
continuous Heisenberg ferromagnet models(the Myrzakulov-IX
equation and its integrable reductions).

The Myrzakulov-IX(M-IX) equation (according to the notations of ref.[8]) looks like
$$ iS_t+\frac{1}{2}[S,M_1S]+A_2S_x+A_1S_y = 0 \eqno(1a)$$
$$ M_2u=\frac{\alpha^{2}}{4i}tr(S[S_y,S_x]) \eqno(1b)$$
where $ \alpha,b,a  $=  consts and
$$
S= \pmatrix{
S_3 & rS^- \cr
rS^+ & -S_3
},\,\,\,\,\,S^{\pm}=S_{1}\pm iS_{2}\,\,\,\,\,\, S^2=I,\,\,\,\,r^{2}=\pm 1
$$
$$ M_1= \alpha ^2\frac{\partial ^2}{\partial y^2}-2\alpha (b-a)\frac{\partial^2}
   {\partial x \partial y}+(a^2-2ab-b)\frac{\partial^2}{\partial x^2}; $$
$$ M_2=\alpha^2\frac{\partial^2}{\partial y^2} -\alpha(2a+1)\frac{\partial^2}
   {\partial x \partial y}+a(a+1)\frac{\partial^2}{\partial x^2},$$
$$
A_1=2i\{(2ab+a+b)u_x-(2b+1)\alpha u_y\}
$$
$$
A_2=2i\{(2ab+a+b)u_y-\alpha^{-1}(2a^2b+a^2+2ab+b)u_x\}.
$$

These set of equations is integrable and admits the some integrable
reductions: the Myrzakulov-VIII equation as b=0 and the Ishimori
equation as $ a=b=-\frac{1}{2} $[8]. In general we have the two
integrable cases: the M-IXA equation as $\alpha^{2} = 1$ and the M-IXB
equation as $\alpha^{2} = -1$. Equation(1) is the (2+1)-dimensional
integrable generalisation  of the Heisenberg ferromagnetic model
$ iS_t=\frac{1}{2}[S,S_{xx}] $ (the isotropic Landau-Lifshitz equation).

The Lax representation of the M-IX equation(1) is given by[8]
$$ \alpha \Phi_y =\frac{1}{2}[S+(2a+1)I]\Phi_x \eqno(2a) $$
$$ \Phi_t=\frac{i}{2}[S+(2b+1)I]\Phi_{xx}+\frac{i}{2}W\Phi_x \eqno(2b) $$
with
$$ W_1=W-W_2=(2b+1)E+(2b-a+\frac{1}{2})SS_x+(2b+1)FS $$
$$ W_2=W-W_1=FI+\frac{1}{2}S_x+ES+\alpha SS_y $$
$$ E = -\frac{i}{2\alpha} u_x,\,\,\,  F = \frac{i}{2}(\frac{(2a+1)u_{x}}{\alpha} -
2u_{y}) $$

Let us now find the equation which is gauge equivalent to the M-IX
equation(1). To this end, we consider the following tranformation
$$ \Phi = g^{-1} \Psi \eqno(3) $$
where $\Phi$ is the matrix solution of linear problem(2), $\Psi$ and
$g$ are a temporally unknown matrix functions. Substituting (3) into (2)
we get
$$ \alpha \Psi_y =\frac{1}{2}[gSg^{-1}+(2a+1)I]\Psi_x
+[\alpha g_{y}-\frac{1}{2}gSg^{-1}g_{x}-\frac{1}{2}(2a+1)g_{x}]g^{-1}\Psi\eqno(4a) $$
$$ \Psi_t=\frac{i}{2}[gSg^{-1}+(2b+1)I]\Psi_{xx}+g\{i[S+(2b+1)I](g^{-1})_{x}
+\frac{i}{2}Wg^{-1}\}\Psi_x +
$$
$$
g\{g_{t}g^{-1} +\frac{i}{2}[S+(2b+1)I](g^{-1}_{xx}
+\frac{i}{2}W(g^{-1})_{x}\}\Psi. \eqno(4b) $$

Now let us choose the unknown function $g$ and $S$ in the form
$$
g= \pmatrix{
f_{1}(1+S_3) & f_{1}rS^- \cr
f_{2}rS^+ & -f_{2}(1+S_3)
},\,\,\,\,\,\,
S = g^{-1}\sigma_{3}g
\eqno(5)
$$
where $f_{j}$ are satisfy the following equations
$$ \alpha (\ln f_{1})_{y}-(a+1)(\ln f_{1})_{x} =\frac{(a+1)(S_{3x}+
S_{3}S_{3x}+S^{-}_{x}S^{+})-\alpha (S_{3y}+S_{3}S_{3y}+
S^{-}_{y}S^{+})}{2(1+S_{3}} \eqno(6a) $$
$$ \alpha (\ln f_{2})_{y}-a(\ln f_{2})_{x} =\frac{a(S_{3x}+
S_{3}S_{3x}+S^{+}_{x}S^{-})-\alpha (S_{3y}+S_{3}S_{3y}+
S^{+}_{y}S^{-})}{2(1+S_{3}} \eqno(6b) $$
It is follows from (4)-(6) that
$$ \alpha g_{y}g^{-1}-B_{1}g_{x}g^{-1}=B_{0}  \eqno(7) $$
where
$$
B_{1}= \pmatrix{
a+1 & 0 \cr
0   & a
},\,\,\,\,\,
B_{0}= \pmatrix{
0   &  q \cr
p   &  0
}.
$$
Here $p, q$ are the some complex functions which are equal
$$ q = \frac{f_{1}\{\alpha[S_{3y}S^{-} -S^{-}_{y}(1+S_{3})]+
(a+1)[S_{3x}S^{-}-S^{-}_{x}(1+S_{3})]\} }{2f_{2}(1+S_{3})} \eqno(8a) $$
$$ p = \frac{f_{2}\{a[S_{3x}S^{+} -S^{-}_{x}(1+S_{3})]+
\alpha[S_{3y}S^{+}-S^{-}_{y}(1+S_{3})]\} }{2f_{1}(1+S_{3})}. \eqno(8b) $$
Hence we obtain
$$ pq = \frac{1}{4}\{\alpha (2a+1){\bf S}_{x}{\bf S}_{y} +
i\alpha{\bf S}({\bf S}_{x}
\wedge {\bf S}_{y}) -a(a+1){\bf S}^{2}_{x}-
\alpha^{2}{\bf S}^{2}_{y}\}. \eqno(9) $$
where ${\bf S}=(S_{1}, S_{2}, S_{3})$ is the three - dimensional
spin(unit) vector.
After these calculations  equations (4) take the forms
$$ \alpha \Psi_y =B_{1}\Psi_x + B_{0}\Psi, \eqno(10a) $$
$$ \Psi_t=iC_{2}\Psi_{xx}+C_{1}\Psi_x+C_{0}\Psi, \eqno(10b) $$
with
$$
C{2}= \pmatrix{
b+1 & 0 \cr
0   & b
},\,\,\,\,\,
C_{1}= \pmatrix{
0   &  iq \cr
ip  &  0
},\,\,\,\,\,
C_{0}= \pmatrix{
c_{11}  &  c_{12} \cr
c_{21}  &  c_{22}
},
$$
$$
c_{12}=i(2b-a+1)q_{x}+i\alpha q_{y},\,\,\,\,\,
c_{21}=i(a-2b)p_{x}-i\alpha p_{y}.
$$
Here $c_{jj}$ satisfy the following system of equations
$$
(a+1) c_{11x}- \alpha c_{11y} =iqp_{x}+p c_{12} -q c_{21},\,
ac_{22x}-\alpha c_{22y}=ipq_{x}-pc_{12}+qc_{21} \eqno (11)
$$

The compatibility condition of equations(10) gives the following
(2+1)-dimensional nonlinear Schr\"odinger type equation
$$
iq_{t}+M_{1}q+vq=0, \eqno(12a)
$$
$$
ip_{t}-M_{1}p-vp=0, \eqno(12b)
$$
$$
M_{2}v = -2M_{1}(pq), \eqno(12c)
$$
which is the Zakharov equation(ZE)[9], where $v=i(c_{11}-c_{22}),\,\,\,\,\,
p=r^{2}q$. Thus we proved that the M-IX equation(1) is
gauge equivalent to the (2+1)-dimensional ZE(12) and vice
versa. In fact the above presented the gauge transformation is reversible.

It is interest note that the M-IX equation(1) admits the some
integrable reductions. Let us now consider these particular integrable
cases.

a) Let $b=0$. Then equations(1) take the form
$$ iS_t=\frac{1}{2}[S_{\xi\xi},S]+iwS_{\xi}   \eqno (13a) $$
$$ w_{\eta}=\frac{1}{4i}tr(S[S_{\xi},S_{\eta}])  \eqno (13b) $$
where
$$
\xi = x+\frac{a+1}{\alpha}y,\,\,\,\,\,
\eta = -x -\frac{a}{\alpha}y,\,\,\,\,\, w=u_{\xi},
$$
which is the M-VIII equation[8]. The gauge equivalent counterpart
of the M-VIII equation(13) we obtain from(12) as $b=0$
$$
iq_{t}+q_{\xi \xi}+vq=0, \eqno(14a)
$$
$$
v_{\eta} = -2r^{2}(\bar q q)_{\xi}, \eqno(14b)
$$
which is the other Zakharov equation[9].

b) Now consider the case: $ a=b=-\frac{1}{2} $. In this case
equations(1) reduces to the well known Ishimori equation
$$ iS_t+\frac{1}{2}[S,(\frac{1}{4}S_{xx}+\alpha^{2}S_{yy})]+
iu_{y}S_x+iu_{x}S_y = 0 \eqno(15a)$$
$$ \alpha^{2}u_{yy} - \frac{1}{4}u_{xx}=
\frac{\alpha^{2}}{4i}tr(S[S_y,S_x]) \eqno(15b)$$

The Ishimori equation (15) is of the great interest since it is the
first example of the integrable spin systems on the plan.
This equation is considered  as a useful laboratory for experimenting with new
theoretical tools ableto handle the specific featuresof soliton models of spin systems
in (2+1)-dimensions. As well known equation(15) allows the rich classes of the
topologically nontrivial and nonequivalent
solutions(solitons, lumps, vortex, dromions and so on) which are classified
by the topological charge
$$
Q = \frac{1}{4\pi}\int \int ^{+\inf}_{-\inf}dxdy{\bf S}({\bf S}_{x}\wedge {\bf S}_{y})  \eqno(16)
$$
The gauge equivalent counterpart of the equation(60) is the Davey-Stewartson
equation
$$ iq_t+\frac{1}{4}q_{xx}+\alpha^{2}q_{yy}+
vq = 0 \eqno(17a)$$
$$ \alpha^{2}v_{yy} - \frac{1}{4}v_{xx}=-2\{\alpha^{2}(pq)_{yy}
 +\frac{1}{4}(pq)_{xx}\} \eqno(17b)$$
that follows from the ZE(12). This fact was for first time
established in [3].

c) Finally let us consider the reduction: $a=-\frac{1}{2}$. Then (1) reduces to the M-XVIII
equation[8]
$$ iS_t+\frac{1}{2}[S,(\frac{1}{4}S_{xx}-\alpha(2b+1)S_{xy}+\alpha^{2}S_{yy})]+
A_{20}S_x+A_{10}S_y = 0 \eqno(18a)$$
$$ \alpha^{2}u_{yy} - \frac{1}{4}u_{xx}=\frac{\alpha^{2}}{4i}tr(S[S_y,S_x]) \eqno(18b)$$
where $A_{j0}=A_{j}$ as $a=-\frac{1}{2}$. The corresponding gauge equivalent equation
obtain from(12) and looks like
$$ iq_t+\frac{1}{4}q_{xx}-\alpha(2b+1)q_{xy}+\alpha^{2}q_{yy}+
vq = 0 \eqno(19a)$$
$$ \alpha^{2}v_{yy} - \frac{1}{4}v_{xx}=-2\{\alpha^{2}(pq)_{yy}-
\alpha (2b+1)(pq)_{xy} +\frac{1}{4}(pq)_{xx}\} \eqno(19b)$$
Note that the Lax representations of equations(13), (15) and (18)
we can get from (2) as $b=0, a=b=-\frac{1}{2}$ and $a=-\frac{1}{2}$
respectively.

In summary, we constructed the gauge equivalent equation to the M-IX
equation which is the Zakharov equation. Also the integrable reductions
of the M-IX equation and their gauge equivalent counterparts are presented.

\end{document}